\documentclass[12pt,letterpaper]{revtex4}
\usepackage{times}
\usepackage{graphicx}
\usepackage{graphics}
\usepackage{dcolumn}
\usepackage{amsmath}
\begin{document}

\title{The absolute frequency of the $^{87}$Sr optical clock transition}

\author{Gretchen K. Campbell, Andrew D. Ludlow, Sebastian Blatt, Jan W. Thomsen, Michael J. Martin, Marcio H. G. de Miranda, Tanya Zelevinsky,  Martin M. Boyd, Jun Ye}

\affiliation{JILA, National Institute of Standards and Technology
and University of Colorado, Department of Physics, University of
Colorado, Boulder, Colorado 80309-0440, USA}

\author{Scott A. Diddams, Thomas P. Heavner, Thomas E. Parker, Steven R. Jefferts}

\affiliation{National Institute of Standards and Technology, Time
and Frequency Division, 325 Broadway, Boulder, Co 80305, USA}
\date{\today}

\begin{abstract}
The absolute frequency of the $^1S_0-$$^3P_0$ clock transition of
$^{87}$Sr has been measured to be 429 228 004 229 873.65 (37) Hz
using lattice-confined atoms, where the fractional uncertainty of
8.6$\times$10$^{-16}$ represents one of the most accurate
measurements of an atomic transition frequency to date. After a
detailed study of systematic effects, which reduced the total
systematic uncertainty of the Sr lattice clock to
1.5$\times$10$^{-16}$, the clock frequency is measured against a
hydrogen maser which is simultaneously calibrated to the US primary
frequency standard, the NIST Cs fountain clock, NIST-F1. The
comparison is made possible using a femtosecond laser based optical
frequency comb to phase coherently connect the optical and microwave
spectral regions and by a 3.5 km fiber transfer scheme to compare
the remotely located clock signals.
\end{abstract}
\maketitle

\section{Introduction}

In recent years optical atomic clocks have made great strides, with
dramatic improvements demonstrated in both stability and accuracy,
and have now surpassed the performance of the best microwave
standards \cite{ludlow08,Rosenband08}. Optical clock candidates are
being investigated by a variety of groups using a number of
different atomic transitions in trapped ions
\cite{Rosenband08,Oskay06,NPLScience04,NRC05,PTBYb05}, trapped
neutral atoms, and freely expanding neutral atoms
\cite{ludlow08,Takamoto05,ludlow06,Targat06,NISTYb2006,Sterr2004,Ertmer05,OatesCa2006}.
As the best optical standards now support an accuracy surpassing
that of the Cs primary standards (3.3 $\times$ 10$^{-16}$)
\cite{NISTCs,SyrteCs,CsPTB2001}, it becomes imperative to directly
compare these optical standards against each other
\cite{Rosenband08,ludlow08} to evaluate them at the lowest possible
level of uncertainties. Nevertheless, it still remains important
that these optical standards are evaluated by the mature primary Cs
standards for multiple reasons.

First, the accuracy of frequency standards is ultimately defined by
the Cs clock under the current realization of the SI-second.
Additionally, over the years a remarkable infrastructure has been
developed to support the transfer of Cs standards for international
intercomparisons, and the primary frequency standards at multiple
national labs all agree within their stated uncertainties
\cite{Parker08}. While fiber networks
\cite{foreman07b,Narbonneau06,Coddington07,Lopez2007} now provide
the most precise frequency distribution links between optical clocks
located near each other (for example within 100~km), for
intercontinental comparisons optical clocks need to be measured
relative to Cs standards. In fact, recent intercomparisons of Sr
clocks among three laboratories at JILA, SYRTE, and University of
Tokyo \cite{Blatt08} have reached an agreement at 1 $\times$
10$^{-15}$, approaching the Cs limit. This has firmly established
the Sr lattice clock standard as the best agreed-upon optical clock
frequency to date, and second only to the Cs standard itself among
all atomic clocks.

Second, an important application of highly accurate atomic clocks is
the test of fundamental laws of nature with high precision. For
example, atomic clocks are placing increasingly tighter constraints
on possible time-dependent variations of fundamental constants such
as the fine-structure constant ($\alpha$) and the electron-proton
mass ratio ($\mu$)
\cite{Marion03,Fischer04,Peik04,Bize05,Peik2006,Lea07,fortier07,Blatt08,Rosenband08}.
These measurements are made by comparing atomic transition
frequencies among a diverse set of atomic species, helping reduce
systematic effects. For example, an optical clock transition
frequency is generally sensitive to variations of $\alpha$, with
different atoms having different sensitivities \cite{Flambaum06}. Sr
in fact has a rather low sensitivity. The Cs standard on the other
hand is based on a hyperfine transition and is sensitive to
variations in both $\alpha$ and $\mu$. Thus measurement of the
frequency ratio of Sr and Cs over the course of a year limits not
only the possible linear drift of these constants but also
constrains possible coupling between fundamental constants and the
gravitational potential, which would signal a violation of local
position invariance \cite{fortier07,Blatt08,Ashby07,Bauch02}.

In recent years, the most accurate absolute frequency measurements
were performed using single trapped ions. These systems benefit from
the insensitivity of the ions to external perturbations, and using
Hg$^+$ ions a frequency uncertainty of 9.1$\times10^{-16}$
\cite{Oskay06} has been achieved. Large ensembles of neutral atoms
offer high measurement signal to noise ratios, however, neutral atom
systems have typically been limited by motional effects. By
confining the atoms in an optical lattice
\cite{Takamoto05,ludlow06,Targat06} these effects are greatly
reduced, as the atoms can be trapped in the Lamb-Dicke regime, where
both Doppler and photon-recoil related effects are suppressed. One
such system is the $^{87}$Sr $(5s^2)^1S_0-$$(5s5p)^3P_0$ transition,
which is currently being pursued by a number of groups worldwide
\cite{boyd07,ludlow08,takamoto06,Baillard07}.

In this paper we report on the absolute frequency measurement of the
$^{87}$Sr $^1S_0-$$^3P_0$ clock transition. The absolute frequency
is measured using a femtosecond laser based octave-spanning optical
frequency comb to compare the $^{87}$Sr optical transition frequency
to a hydrogen maser, which is simultaneously calibrated to the NIST
fountain primary frequency standard, NIST-F1. To remotely link the
Sr standard, which is located at JILA on the University of Colorado
campus, to the NIST-F1 Cs clock, located at the NIST Boulder
laboratories, a 3.5 km optical fiber link is used to transfer the
H-maser reference signal \cite{Ye03,foreman07}. In addition to
demonstrating one of the most accurate measurements of an optical
transition frequency to date, the agreement of this result with
previous measurements both at JILA and around the world demonstrates
the robustness and reproducibility of strontium as a frequency
standard, and as a future candidate for the possible redefinition of
the SI second.

\section{Experimental Setup}
\begin{figure}
\centering{
\includegraphics [width=\columnwidth]{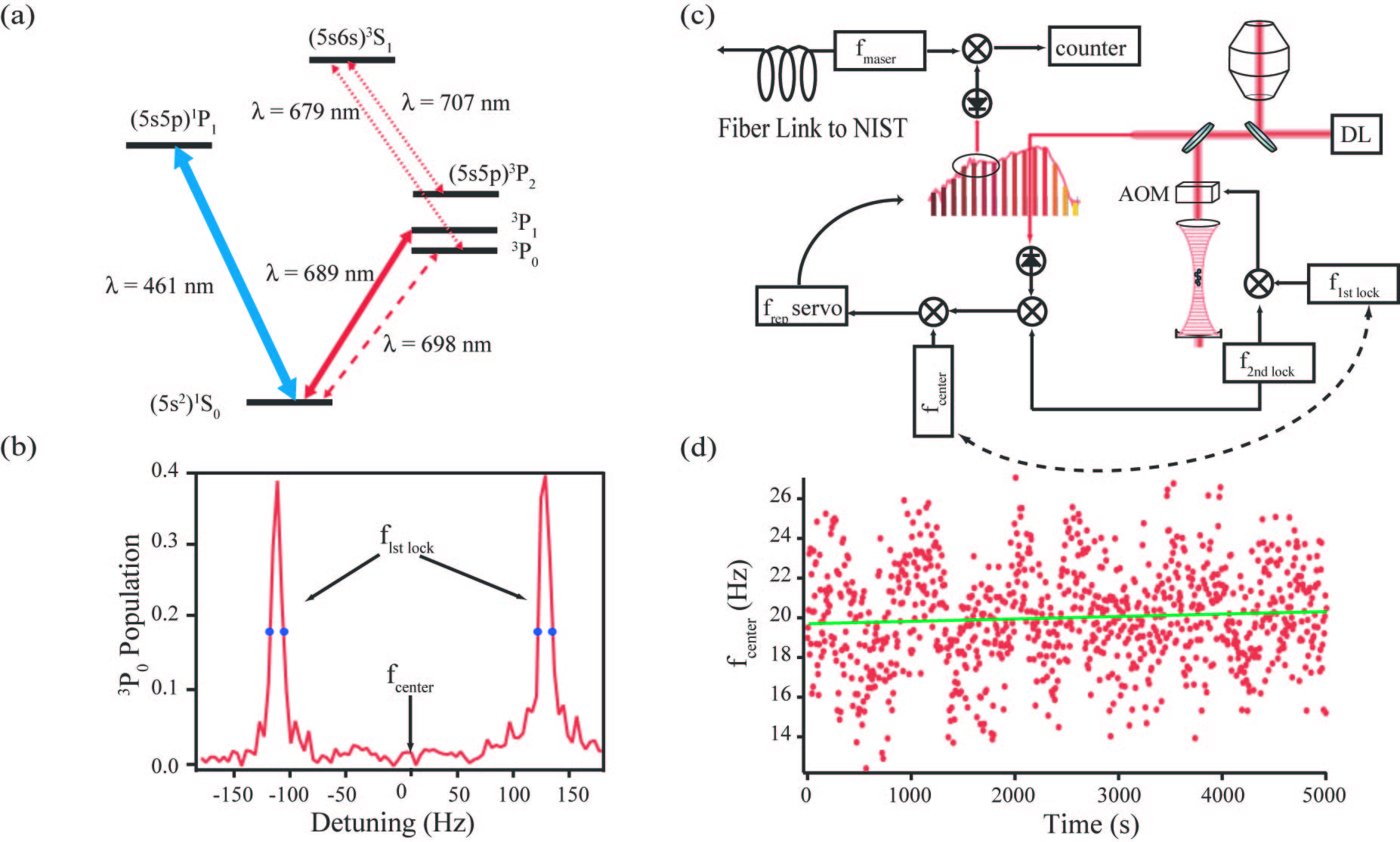}
\caption{\label{fig:figone} Experimental Setup. (a) Relevant energy
levels for $^{87}$Sr used for the optical lattice clock. Transitions
at 461 nm and 689 nm are used in two-stage cooling and trapping of
the Sr atoms. The clock transition is at 698 nm. Lasers at 679 and
707 nm provide necessary repumping from metastable states. (b) To
operate the clock, ultracold $^{87}$Sr atoms are first optically
pumped to the $|F = 9/2,m_f = \pm9/2\rangle$ states. The clock
center frequency ($f_{center}$) is found by locking the probe laser
frequency to both peaks successively and taking their average. The
laser is locked to the center of each transition by sampling their
FWHM as illustrated in the figure by dots ($f_{1st}$ $_{lock}$). (c)
Schematic of the setup used for locking the optical local oscillator
to the $^{87}$Sr transition. The clock transition is probed using a
diode laser (DL) at $\lambda$ = 698 nm which is locked to an
ultrastable, high finesse optical cavity.  The laser beam is used to
interrogate the Sr atoms and is transferred to the atoms using an
optical fiber with active fiber noise cancelation. To steer the
frequency of the laser for the lock to the Sr resonance, an
acousto-optic modulator (AOM) is used to introduce a frequency
offset between the cavity and the atoms. The frequency offset is
steered to the lock points ($f_{1st}$ $_{lock}$). The frequency
offset also includes a linear feedback value ($f_{2nd}$ $_{lock}$)
to compensate for the linear drift of the high finesse cavity. The
cavity-stabilized clock laser is also used to phase-lock a
self-referenced octave-spanning optical frequency comb, in
combination with atomic resonance information contained in
$f_{center}$ and $f_{2nd}$ $_{lock}$. The Sr-referenced repetition
frequency of the comb ($f_{rep}$) is then counted relative to a
H-maser located at NIST ($f_{maser}$). (d) Sample data showing the
in-loop atom lock for 5000 s of data taken during the measurement of
the absolute frequency. The fit gives a residual linear fractional
frequency drift of $<$ 2$\times10^{-19}$/s.}}
\end{figure}

The frequency standard uses lattice-confined $^{87}$Sr atoms with
nuclear spin $I$ = 9/2. Although the $^{87}$Sr apparatus has been
previously described elsewhere \cite{boyd07,ludlow06}, here we
summarize the experimental details most relevant to this work. To
measure the frequency of the clock transition, $^{87}$Sr atoms are
first trapped and cooled to mK temperatures in a magneto-optical
trap (MOT) operated on the $^1S_0-$$^1P_1$ strong 461 nm cycling
transition (see Fig.~\ref{fig:figone}a for a diagram of relevant
energy levels). The atoms are then transferred to a second stage 689
nm MOT for further cooling. This dual-frequency MOT uses narrow line
cooling \cite{Mukaiyama03,Loftus2}, resulting in final temperatures
of $\sim$1 $\mu$K. During the cooling process, a one-dimensional
optical lattice is superimposed in the nearly vertical direction.
After the second MOT stage, the MOT optical beams and the
inhomogeneous magnetic field are turned off, leaving $\sim$10$^4$
atoms at 2.5 $\mu$K trapped in the optical lattice. The optical
lattice is created using a retro-reflected laser beam and is
operated near a laser frequency where the polarizability of the
$^1S_0$ and $^3P_0$ states are identical for the lattice field
\cite{Katori03,brusch06}. For this work, the lattice is operated at
a trap depth of $U_T =$ 35$E_{rec}$, where $E_{rec}=\hbar^2 k^2/2m$
is the lattice photon recoil energy and $k=2\pi/\lambda$ is the
wavevector of the lattice light.  At this lattice depth the atoms
are longitudinally confined in the Lamb-Dicke regime and in the
resolved sideband limit \cite{Leibfried03}. Spectroscopy is
performed by aligning the probe laser precisely along the axis of
the lattice standing wave, and the atoms are probed free of recoil
or motional effects. The vertical orientation of the lattice breaks
the energy degeneracy between lattice sites, strongly prohibiting
atomic tunneling \cite{Lemonde3}.

Before performing spectroscopy, the atoms are first optically pumped
to the stretched $|$$F=$ 9/2, $m_F=\pm$9/2$\rangle$ states with the
use of a weak optical beam resonant with the $^1S_0$($F=$
9/2)-$^3P_1$ ($F=$ 7/2) transition. Here $\vec{F}$ = $\vec{I}$ +
$\vec{J}$ is the total angular momentum, with $\vec{I}$ the nuclear
spin and $\vec{J}$ the total electron angular momentum. The beam
used for optical pumping is aligned collinear with the lattice, and
is linearly polarized along the lattice polarization axis. The
optical pumping is performed with a small magnetic bias field
($\sim$3 $\mu$T), which is also oriented along the lattice
polarization. After optical pumping, spectroscopy is performed on
the $^1S_0$-$^3P_0$ clock transition from the two spin sublevels.
The clock transition, which has a theoretical natural linewidth of
$\sim$1 mHz \cite{boyd07b,Santra04,Porsev04,Katori03}, is
interrogated using a diode laser at 698 nm, which is prestabilized
by locking it to a high-finesse ultrastable cavity, resulting in a
laser optical linewidth below 1 Hz \cite{Ludlow07}. The probe beam
is coaligned and copolarized with the optical lattice. To ensure
that the stretched states are well resolved, the spectroscopy is
performed under a magnetic bias field of 25 $\mu$T, which results in
a $\sim$250 Hz separation between the two $\pi$-transitions excited
during the spectroscopy.

Spectroscopy is performed using an 80-ms Rabi pulse, which when on
resonance transfers a fraction of the atoms into the $^3P_0$ state.
After applying the clock pulse, atoms remaining in the $^1S_0$
ground state are detected by measuring fluorescence on the strong
$^1S_0-$$^1P_1$ transition. The length of the pulse is long enough
to measure both the population in the $^1S_0$ state as well as to
heat these atoms out of the trap. The population in the $^3P_0$
state is then measured by first pumping the atoms back to the
$^1S_0$ state through the intermediate $(5s5p)^3P_0-$$(5s6s)^3S_1$
and $(5s5p)^3P_2-$$(5s6s)^3S_1$ states and then by again measuring
the fluorescence on the $^1S_0-$$^1P_1$ transition. Combining these
two measurements gives a normalized excitation fraction insensitive
to atomic number fluctuations from shot to shot. A typical spectrum
is shown in Fig.~\ref{fig:figone}b. The Fourier-limited linewidth of
the transition is 10 Hz, much less than the 250 Hz separation
between the peaks, which makes the lines well resolved and also
reduces potential line pulling effects due to any residual
population left in other spin states by imperfect optical pumping.
We note that while our optical local oscillator supports recovery of
$<$ 2 Hz spectroscopic linewidths \cite{Boyd2006}, we find it more
robust to run the clock transition with a 10 Hz Fourier-limited
spectral linewidth.

To stabilize the optical local oscillator used for spectroscopy to
the atomic transition, we use both stretched states. Using two
time-multiplexed independent servos, we lock the laser frequency to
the center of each transition. This is done by sampling the full
width half maximum (FWHM) of each transition (labeled $f_{1st}$
$_{lock}$ in Fig.~\ref{fig:figone}b). The average of the two line
centers gives the center frequency of the clock transition. The
cavity-stabilized local oscillator, in combination with this
frequency $f_{center}$, is in turn used to phase-lock a
self-referenced octave-spanning optical frequency comb. The
Sr-referenced repetition frequency of the comb is then counted with
a H-maser located at NIST.  A schematic of this locking setup is
shown in Fig.~\ref{fig:figone}c. Determination of the center
frequency requires four experimental cycles, two for each of the
$m_F=\pm$9/2 transitions since a new atomic sample is reloaded for
each lock point. The length of each experimental cycle is $\approx$
1.1s. After first probing the $\pi$ transition for the $m_F= -$9/2
transition, we then probe the transition for the $m_F= +$9/2 state.
The digital servo operates via standard modulation techniques. A
linear feedback is also implemented to compensate for the drift of
the high-finesse cavity used to prestabilize the clock laser. A
second integration stage in the laser-atom feedback loop is used to
calculate this feedback value (labeled $f_{2nd}$ $_{lock}$) in
Fig.~\ref{fig:figone}c. As shown in Fig.~\ref{fig:figone}d, using
this approach limits the residual drifts compensated for by the
first servo integrator to typically $<$1 mHz/s.

\section{Systematic shifts of the Strontium Clock}

We have recently evaluated the systematic shifts of the strontium
clock at the $1\times10^{-16}$ level \cite{ludlow08}, and in Table
\ref{tbl:critical} the important systematic shifts to the absolute
frequency are shown. Although a detailed description can be found in
\cite{ludlow08}, here we summarize these shifts.  The evaluation of
the systematic uncertainty is performed using the remotely located
calcium optical standard at NIST \cite{Oates06}, which is linked to
JILA via a phase coherent optical fiber link \cite{foreman07b}. The
Sr-Ca comparison has a 1 s stability of 2$\times10^{-15}$ which
averages down to below 3$\times10^{-16}$ after 200 s. To measure the
Sr systematics an interleaved scheme is used where the Sr parameter
of interest is varied between two different settings every 100 s,
while the Ca standard remains locked. Pairs of such data are then
used to determine the frequency shift, and many pairs are averaged
in order to reach below the $10^{-16}$ level.
\renewcommand{\arraystretch}{.6}
\begin{table}[h!]
\begin{center}
\begin{tabular}[t]{l c c}
\hline\hline
Contributor & Correction (10$^{-16}$)     & Uncertainty (10$^{-16}$)\\
\hline
Lattice Stark (scalar/tensor) & -6.5 & 0.5 \\
Hyperpolarizability (lattice) & 0.1 & 0.1 \\
BBR Stark & 54.0 & 1.0 \\
AC Stark (probe beam) & 0.15 & 0.1 \\
1$^{st}$ order Zeeman & 0.2 & 0.2 \\
2$^{nd}$ order Zeeman & 0.36 & 0.04 \\
Density & 3.8 & 0.5 \\
Line pulling & 0 & 0.2 \\
Servo error & 0 & 0.5 \\
2$^{nd}$ order Doppler & 0 &$\ll$0.01 \\
\hline  Sr Systematics Total& 52.11 & 1.36\\[1ex]

Maser calibration &-4393.7&8.5 \\
Gravitational shift & 12.5 & 1.0 \\
\hline Total & -4329.1 & 8.66 \\
\hline\hline $\nu_{Sr}-\nu_0$ & 73.65 Hz & .37 Hz \\
\hline\hline
\end{tabular}
\renewcommand{\arraystretch}{1}
\caption{ Frequency corrections and their associated uncertainties
for the clock transition in units of 10$^{-16}$ fractional
frequency, and with $\nu_0=$429 228 004 229 800 Hz. The maser
correction uncertainty includes both Sr/H-Maser comparison, as well
as the Cs clock uncertainty.} \label{tbl:critical}
\end{center}
\end{table}

As shown in Table \ref{tbl:critical}, besides the correction that
arises from the maser calibration, the dominant shift for the Sr
clock is the black-body radiation (BBR)-induced shift. To determine
this shift, the temperature of the Sr vacuum chamber is continuously
monitored during the course of the absolute frequency measurement at
four separate locations. During the measurement the average
temperature of the chamber is 295(1) K, and the corresponding BBR
effect:
\begin{equation}
\delta\nu_{BBR}=-2.354(32)\left(\frac{T}{300~\mathrm{K}}\right)^4
\mathrm{Hz}
\end{equation}
gives a frequency shift of 54(1)$\times10^{-16}$. Higher-order
multipoles are suppressed by $\alpha^2$ and are negligible at this
level. The given uncertainty in the BBR shift includes the error due
to the chamber temperature, as well as the theoretical uncertainty
in the polarizability \cite{porsev06}.

For the duration of the experiment, the lattice laser is phase
locked to the same optical frequency comb used to count the Sr beat,
and the wavelength is simultaneously monitored on a wavemeter to
ensure it does not mode-hop. The lattice is operated at a frequency
of 368554.36(3) GHz \cite{ludlow06,ludlow08}, slightly away from the
state-insensitive lattice frequency. Including nuclear spin effects,
the light shift due to the linearly polarized lattice can be
expressed as \cite{boyd07b}:
\begin{equation}
\delta\nu_S\approx-(\Delta\kappa^S-\Delta\kappa^T
F(F+1))\frac{U_T}{E_{rec}}-(\Delta\kappa^Vm_F\xi\cos(\varphi)+\Delta\kappa^T3m_F^2)\frac{U_T}{E_{rec}}
\end{equation}
where $\Delta\kappa^{S,T,V}$ is the frequency shift coefficient due
to the differential polarizability (scalar, tensor, and vector)
between the ground and excited clock states, $\xi$ is the degree of
ellipticity of the beam (with $\xi$ = 0 for $\pi$-polarized light),
and $\varphi$ is the angle between the lattice propagation direction
and the bias magnetic field ($\simeq \pi/2$ for our setup). For the
linearly polarized lattice configuration in our setup, the vector
light shift is minimized; furthermore since the Sr clock is operated
using both the $m_F=\pm$ 9/2 states, the antisymmetric $m_F$
dependence averages away this vector shift. The effect of the tensor
light shift for a given $|m_F|$ state introduces a polarization
dependent offset to the state-insensitive lattice frequency.
Experimentally, this Stark cancelation frequency for the
$m_F=\pm9/2$ state has been determined to be 368554.68(17) GHz
\cite{ludlow08} leading to a shift of -6.5(5)$\times10^{-16}$ to the
absolute clock frequency for the lattice depth and frequency used
during this measurement. For our operating conditions,
hyperpolarizability effects are more than an order of magnitude
smaller \cite{brusch06, Targat06}.  The ground and excited clock
states have different polarizabilities at the clock transition
frequency, and imperfect alignment between the clock laser and the
lattice beam can lead to inhomogeneous Rabi frequencies in the
transverse direction requiring an increase in the clock transition
probe power. However, given the small saturation intensity of the
clock probe beam, the ac Stark shift introduced by the clock laser
during spectroscopy is small, and has been experimentally measured
to be 0.15(10)$\times10^{-16}$. Stark shifts from laser beams not
used during spectroscopy, for example those used for cooling and
trapping during the MOT phase and for fluorescence detection after
spectroscopy, are eliminated through the use of acousto-optic
modulators (AOM) in series with mechanical shutters which block
these beams during spectroscopy. In addition, the vacuum chamber is
covered with an opaque cloth to prevent any stray light from
entering the chamber.

For each experimental cycle sequence, the total atom number is
recorded, allowing a point-for-point correction of the density
shift. The value given in Table \ref{tbl:critical} is the average
density correction. At the FWHM of the spectroscopic signal where
the probe laser is locked, the excitation fraction in each stretched
state is 15(2)\%. This excitation fraction and our operating density
of $\simeq$ 4$\times$10$^{10}$cm$^{-3}$ gives a frequency correction
of 3.8(5)$\times$ $10^{-16}$ \cite{ludlow08,ludlow08b}.

The Zeeman shift of the transition frequency is given in Hz by
\cite{boyd07b}
\begin{equation}
\label{eq:zeeman} \delta\nu_B\approx-\delta g m_F\mu_0 B+\beta B^2
\approx-1.084(4)\times 10^{6}m_FB- 5.8(8)\times10^{-8} B^2
\end{equation}
where $\mu_0$ = $\mu_B/h$, with $\mu_B$ the Bohr magneton, and
$\delta g$ the differential Land\'{e} g-factor between the ground
and excited clock states. The $1^{st}$-order Zeeman shift is
experimentally measured in \cite{boyd07b}, and the $2^{nd}$-order
Zeeman shift is experimentally measured in \cite{ludlow08},
consistent with other measurements \cite{Baillard07b}. By measuring
the average frequency of the stretched states at a small bias field,
the $1^{st}$ order Zeeman effect is averaged away due to the
opposite linear dependence of the shift on the $m_F=\pm 9/2$ states,
and the experimentally measured value for the shift is consistent
with zero. The bias field of 25 $\mu$T used during spectroscopy is
large enough such that the spin states are well resolved, reducing
line pulling effects due to residual populations in other states,
yet small enough such that the $2^{nd}$ order Zeeman shift is
negligible, with a value of 0.36(4)$\times$10$^{-17}$ for our bias
field.

By operating in the Lamb-Dicke regime, $1^{st}$ order Doppler shifts
are minimized. However, driven motion can also cause frequency
shifts due to shaking of the lattice beams, or due to relative
motion between the lattice and the probe beam. To minimize
vibrations, the optics table is floated using standard pneumatic
compressed air legs, and we estimate the effect of $1^{st}$ order
Doppler shifts to be below $10^{-18}$. Switching magnetic fields can
also induce vibrations, however, our quadrupole trap is switched off
more than 100 ms before spectroscopy. Furthermore, $2^{nd}$-order
Doppler effects from residual thermal motion are negligible
($<10^{-18}$), given the T = 2.5 $\mu$K temperature of the trapped
Sr atoms.

The digital servos used to steer the spectroscopy laser to the
atomic transition are another potential source of frequency offsets.
The dominant cause of servo error is insufficient feedback gain to
compensate for the linear drift of the high-finesse reference
cavity. The second integration step as described in the experimental
setup section reduces this effect. By analyzing our servo record we
conservatively estimate this effect to be $<5\times10^{-17}$. In
conclusion, with the exception of the BBR-induced shift, all of the
systematics discussed above are limited only by statistical
uncertainty.

\section{Fiber transfer between JILA and NIST}

\begin{figure}
\centering{
\includegraphics[hiresbb=true, width=8 cm]{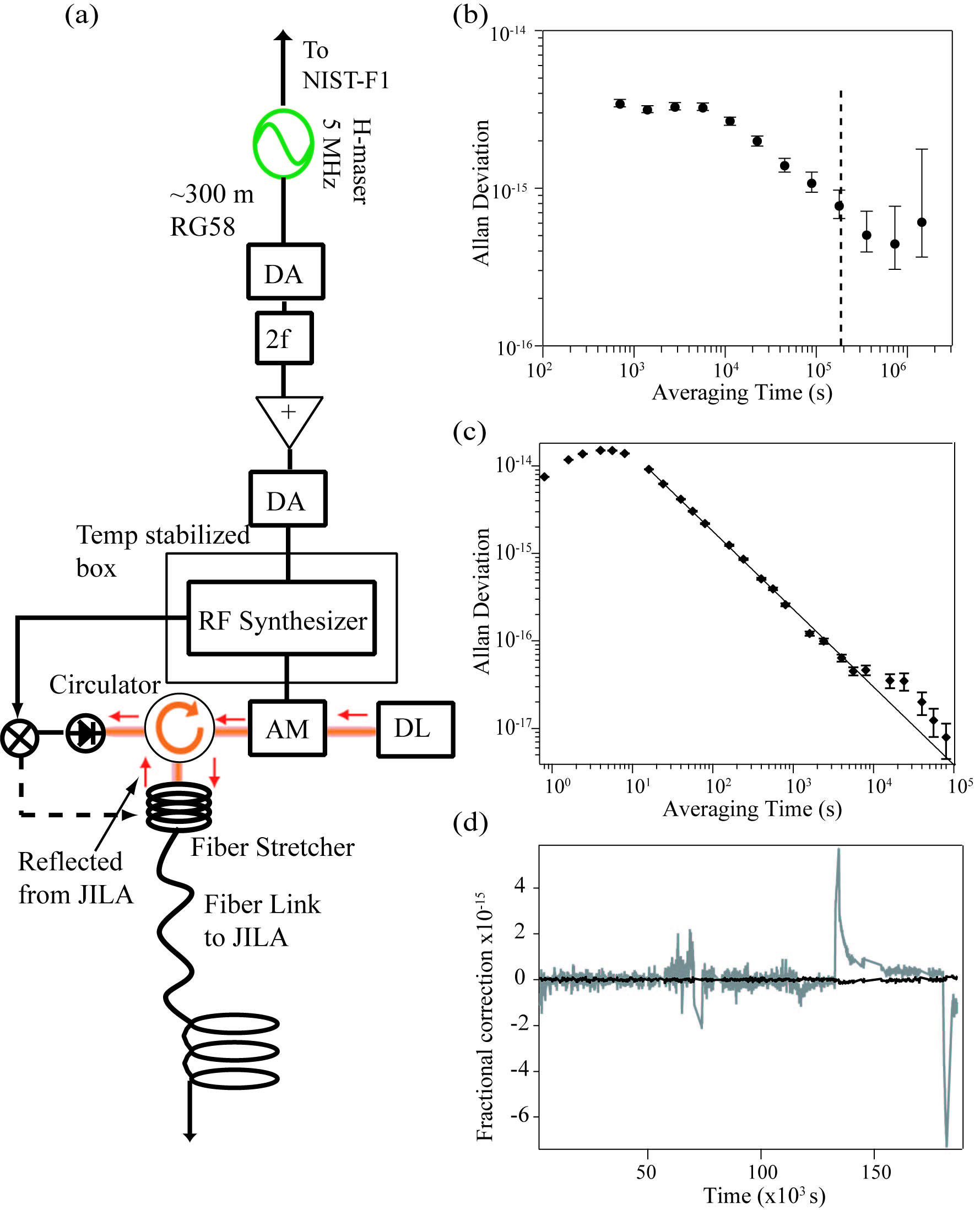}
\caption{\label{fig:figNIST} Clock signal transfer between NIST and
JILA. (a) Schematic of the setup used to transfer the hydrogen maser
signal from NIST to JILA. A 5 MHz signal from the H-maser, which is
simultaneously counted against the NIST-F1 Cs standard, is
distributed through a $\sim$300 m cable to a distribution amplifier
(DA). After the distribution amplifier it is actively frequency
doubled (2f) and sent through a second distribution amplifier. The
resulting 10 MHz signal is used to reference an RF Synthesizer
operating at $\sim$950 MHz. The synthesizer in turn modulates the
amplitude (AM) of a 1320 nm laser (DL) which is transferred to JILA
through a 3.5 km fiber link. Noise from the fiber link is canceled
with a fiber stretcher to actively stabilize the microwave phase
using a retroreflection of the beam sent back from JILA
\cite{foreman07}. (b) Typical Allan deviation of the H-maser used
for the Sr absolute frequency measurement, with the total duration
of the measurement represented by a dotted line. (c) Out of loop
measurement of the stability of the microwave electronics used for
transfer and fiber noise cancelation. The fit to the line gives a
1-s Allan deviation of 1.08(1)$\times$$10^{-13}$ with a slope of
-0.889(4). The bump at 10000 s is indicative of temperature
fluctuations in the distribution room during the out of loop
measurement. The bump at 10 s is due to low pass filtering of the
phase measurement. (d) Frequency correction due to temperature
fluctuations in the distribution room (grey curve) and fluctuations
in the temperature-stabilized box used to house the RF synthesizer
(black curve) during the course of the measurement.}}
\end{figure}

The strontium experiment is located at JILA, on the University of
Colorado, Boulder campus and is linked to NIST Boulder Laboratories
by a 3.5 km optical fiber network. To measure the absolute frequency
of the transition, the tenth harmonic of the repetition rate of the
frequency comb (which is phase-locked to the Sr clock laser and
located at JILA), is beat against a $\sim$950 MHz signal originating
from NIST. A schematic of the transfer scheme from NIST is shown in
Fig.~\ref{fig:figNIST}. A 5 MHz signal from a hydrogen maser is
distributed along a $\sim$300 m cable to a distribution room where
it is first frequency doubled and then used as a reference to
stabilize an RF synthesizer operating near 950 MHz. In
Fig.~\ref{fig:figNIST}b, the stability of the H-maser as measured by
Cs is shown, where the total length of the Sr absolute frequency
measurement is indicated with a dotted line. The RF signal generated
by the synthesizer is then used to modulate the amplitude of a 1320
nm diode laser which is then transferred to JILA via the 3.5 km
optical fiber link between the two labs \cite{Ye03,foreman07}. The
microwave phase of the fiber link is actively stabilized using a
fiber stretcher to control the group delay between NIST and JILA
\cite{foreman07}. The limited dynamic range of the fiber stretcher
necessitates a periodic change of the transfer frequency to relock
the fiber transfer system for our 3.5 km link. Typically the dynamic
range is sufficient to stabilize drifts for roughly 30-60 minute
intervals, after which it must be unlocked and reset, leading to a
dead time in the measurement of $\sim$1 minute.

The transfer of the microwave signal between the NIST H-maser and
JILA can potentially introduce a number of systematic frequency
shifts and uncertainties. The majority of these arise from
temperature-driven fluctuations during the course of the
measurement. The microwave signal is transferred between the maser
and the RF synthesizer using 300 m of cable after which it goes
through a series of distribution amplifiers and a frequency doubler.
All of the microwave electronics, as well as the cable used to
transfer the signal, are sensitive to temperature-driven phase
excursions. In order to correct for these effects, the temperature
in the room is monitored continuously during the course of the
experiment. In addition, the RF synthesizer is placed in a
temperature-stabilized, thermally insulated box, and the temperature
in the box is also continuously monitored.  The temperature
coefficient of the synthesizer is independently measured by applying
a temperature ramp to the box while counting the frequency of the
synthesizer relative to a second frequency stable synthesizer. The
synthesizer is found to have a temperature coefficient of -3.6 ps/K,
corresponding to a fractional frequency change of -1$\times10^{-15}$
for a temperature ramp of 1 K/h. This temperature coefficient is
used to make rolling frequency corrections during the absolute
frequency measurement.

To test the performance of the microwave electronics used for the
modulation of the transfer laser as well as the fiber noise
cancelation, an out-of-loop measurement is performed by detecting
the heterodyne beat between the resulting transfer signal and the RF
synthesizer. In Fig.~\ref{fig:figNIST}c the Allan deviation is shown
for this measurement, demonstrating that the fractional frequency
instability due to these transfer components is 1$\times10^{-14}$
from 1-10 s, and averages down to $<10^{-17}$ at $10^5$ s. By
correlating the out-of-loop measurement with temperature
fluctuations in the distribution room during the course of the
measurement, a temperature coefficient of 4.4 $\times
10^{-16}$/(K/h) is found for the microwave electronics.

The distribution amplifiers, frequency doubler, and cable used to
transfer the signal within the distribution room are also tested by
comparing the 10 MHz signal used to stabilize the synthesizer with a
signal split off before the distribution amplifiers. This
measurement determined a coefficient of $\sim7\times10^{-16}$/(K/h).

The absolute frequency measurement, as discussed below in Section V,
is recorded over 50 continuous hours.  During the course of the
absolute frequency measurement the insulated box used to house the
RF synthesizer maintained an average temperature of 292.2(2) K, with
a maximum slope of $<0.1$ K/h. In Fig.~\ref{fig:figNIST}d the black
trace shows the resulting fractional frequency correction to the Sr
frequency due to temperature-driven frequency fluctuations of the RF
synthesizer. The total fractional correction to the Sr frequency due
to the microwave electronics during the measurement, as well as the
distribution amplifiers and cables, is shown by the grey trace in
Fig.~\ref{fig:figNIST}d. During hour 37 (133$\times10^3$ s) of the
measurement, a temperature ramp began in the distribution room,
leading to a large slope in the temperature during hour 37 and
during hour 50 (180$\times 10^3$ s) as the temperature restabilized.
However, this transient affected only a small fraction of the data.
Using the measured temperature coefficients, a rolling correction is
made to all of the measured frequencies. The average frequency
correction during the course of the measurement is
$9\times10^{-17}$, with an uncertainty of $1\times10^{-17}$. These
corrections do not influence the statistics of the final absolute
frequency measurement. In other words, the final mean frequency and
standard error are the same with and without these corrections.

\section{Frequency measurement results}

As shown in Fig.~\ref{fig:figNIST}b, the Allan deviation of the
H-maser averages down as $\sim$ 3$\times10^{-13}/\sqrt{\tau}$. To
measure the Sr absolute frequency to below $10^{-15}$, the
measurement is performed for 50 continuous hours. The largest
frequency correction to the measured $^{87}$Sr frequency is the
calibration offset of the H-maser frequency. The H-maser is
simultaneously counted against the Cs standard during the duration
of the measurement and the resulting frequency correction to the
$^{87}$Sr/H-maser comparison is -439.37(85)$\times 10^{-15}$, where
the uncertainty includes both an uncertainty of 0.6$\times 10^{-15}$
due to the Cs standard, as well as an uncertainty of 0.6$\times
10^{-15}$ from dead time in the Sr/H-maser measurement. An
additional frequency correction is the gravitational shift due to
the difference in elevation between the Cs laboratory at NIST and
the Sr laboratory at JILA. The difference in elevation between the
two labs, which has been determined using GPS receivers located in
each building to be 11.3(2) m, gives a frequency shift of
12.5(1.0)$\times10^{-16}$, where the given uncertainty includes the
uncertainty in the elevation as well as the uncertainty due to the
geoid correction \cite{geoid}.

\begin{figure}
\centering{
\includegraphics[hiresbb=true]{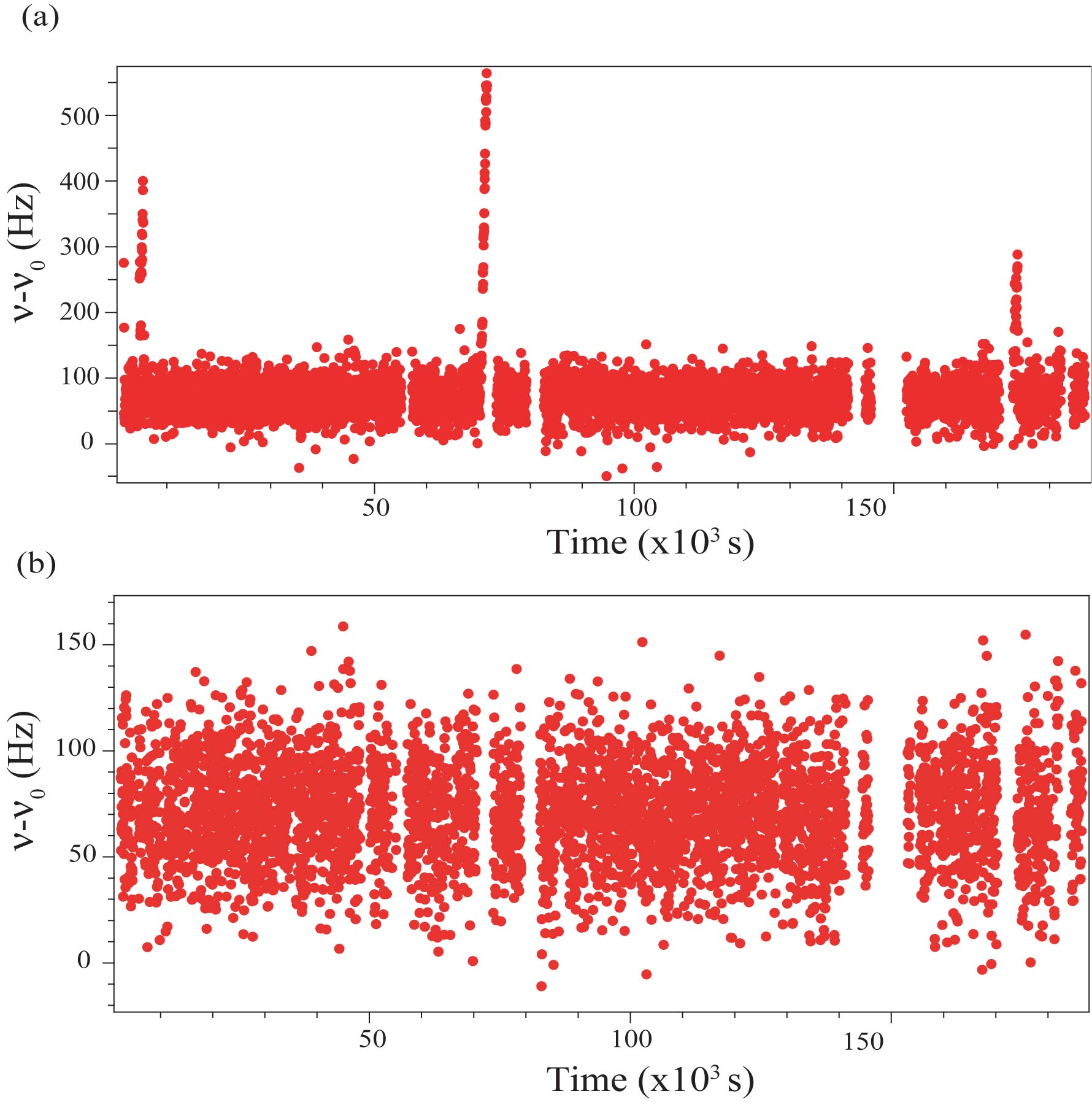}
\caption{\label{fig:figtwo} Absolute frequency measurements of the
$^1$S$_0-^3$P$_0$ clock transition. (a) Counting record showing all
of the data taken over a 50 hour period. Each point corresponds to a
30 s average, and the overall offset is $\nu_0$ = 429 228 004 229
800 Hz. (b) The counting record after removing points where the
system is not locked. The mean value is 70.88(35) Hz.}}
\end{figure}

\begin{figure}
\centering{
\includegraphics[hiresbb=true, width=\columnwidth]{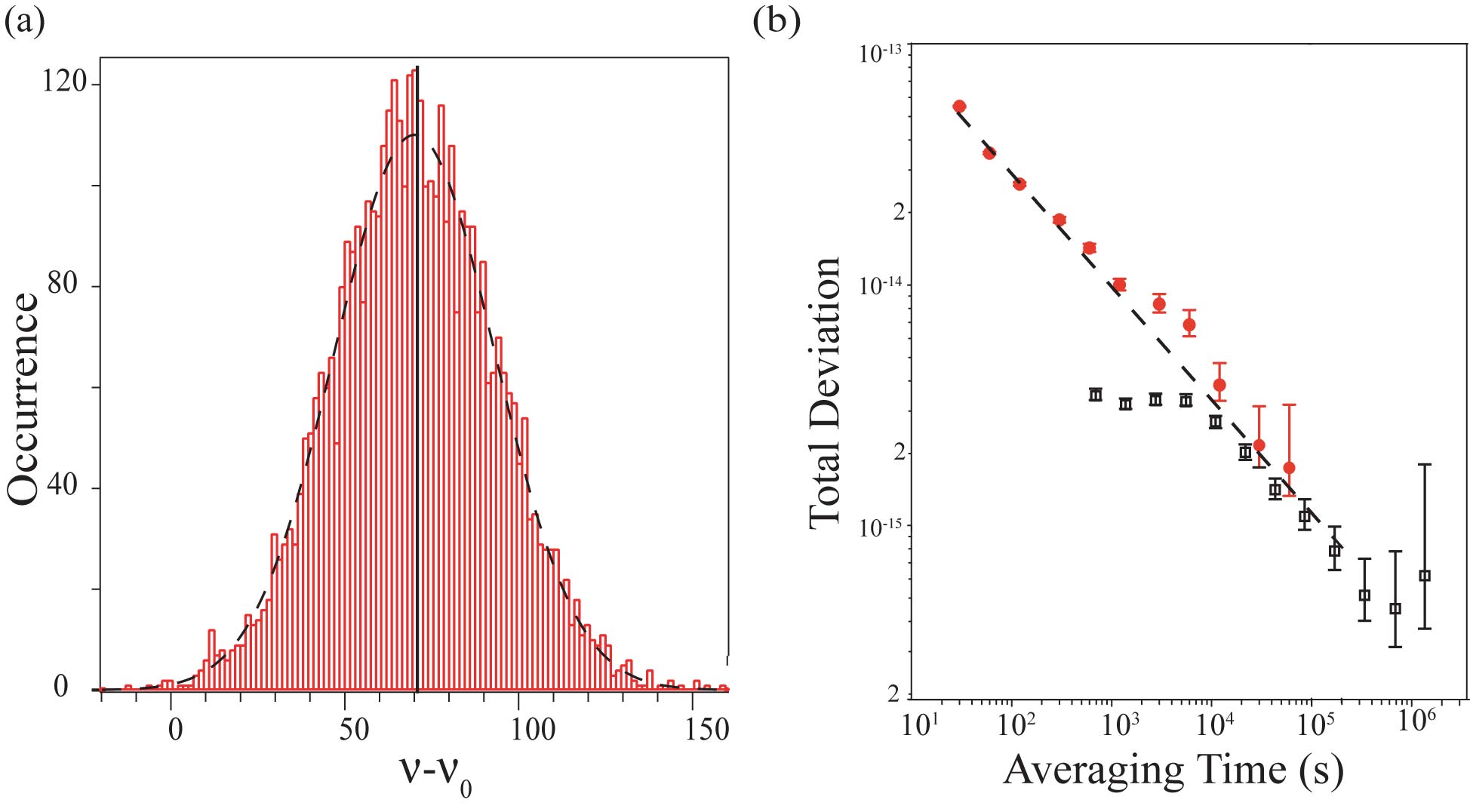}
\caption{\label{fig:figthree} (a) Histogram of the frequency
measurements shown in Fig.~\ref{fig:figtwo}, including the maser
correction. The dashed line is a gaussian fit to the data, the mean
frequency is 70.88 Hz and is indicated by the black line. (b) Total
Deviation of the frequency measurement for the Sr/H-maser comparison
(circles), and the H-maser/Cs comparison (squares). The dotted line
shows a fit of the Sr deviation to $a\tau^{-b}$, where $a$ =
2.64(8)$\times10^{-13}$ and b = 0.48(1), and the dotted line extends
out to the full measurement time. For averaging times $\tau>10^4$ s,
the maser noise dominates both the Cs/H-maser and the Sr/H-maser
measurement, and hence the maser uncertainty (6$\times10^{-16}$, as
described in the text) needs to be counted only once in the final
Sr/Cs measurement uncertainty budget.}}
\end{figure}

In Fig.~\ref{fig:figtwo}, the 50 hour counting record is shown for
the Sr frequency, with a 30 s gate time. The frequency shown
includes only the correction due to the maser, and is plotted with
an offset frequency of $\nu_0$ = 429 228 004 229 800 Hz. The
frequency excursions and gaps seen in Fig.~\ref{fig:figtwo}a occur
when the Sr system is unlocked, which happens when either the
frequency comb comes unlocked, or when the probe laser is not locked
to the atomic signal. During the course of the measurement the
lattice intensity and frequency, all laser locks, and the
temperature at both JILA and in the distribution room in NIST are
continuously monitored and recorded. In Fig.~\ref{fig:figtwo}b, data
corresponding to times when any lasers are unlocked, including times
when the spectroscopy laser is not locked to the atoms, and times
when the lattice laser intensity or frequency is incorrect, have
been removed. In Fig.~\ref{fig:figthree}a, a histogram of this final
counting record is shown, demonstrating the gaussian statistics of
the measurement. The mean value (relative to $\nu_0$) of the
measured frequency is 70.88(35) Hz. In Fig.~\ref{fig:figthree}b the
Total deviation of the frequency measurement is shown. The Total
deviation \cite{Greenhall99} is similar to the Allan deviation,
however it is better at predicting the long-term fractional
frequency instability. The 1-s stability of the H-maser used for the
measurement is 1.5$\times10^{-13}$. However, from a fit to the Total
deviation, we find a 1-s stability of $\sigma_{1s}$ =
2.64(8)$\times10^{-13}$ for the counting record, which is limited by
counter noise, and averages down as $\sigma_{1s}$ /$\tau^{0.48(1)}$.
Extrapolating to the full length of the data set (excluding dead
time in the measurement) gives a statistical uncertainty of 8$\times
10^{-16}$. The frequency uncertainty of Sr is low enough such that
this uncertainty is dominated by the performance of the maser (see
Fig.~\ref{fig:figthree}b), which is included in the maser
calibration uncertainty given in Table \ref{tbl:critical}.

\begin{figure} \centering{
\includegraphics[hiresbb=true, width=\columnwidth]{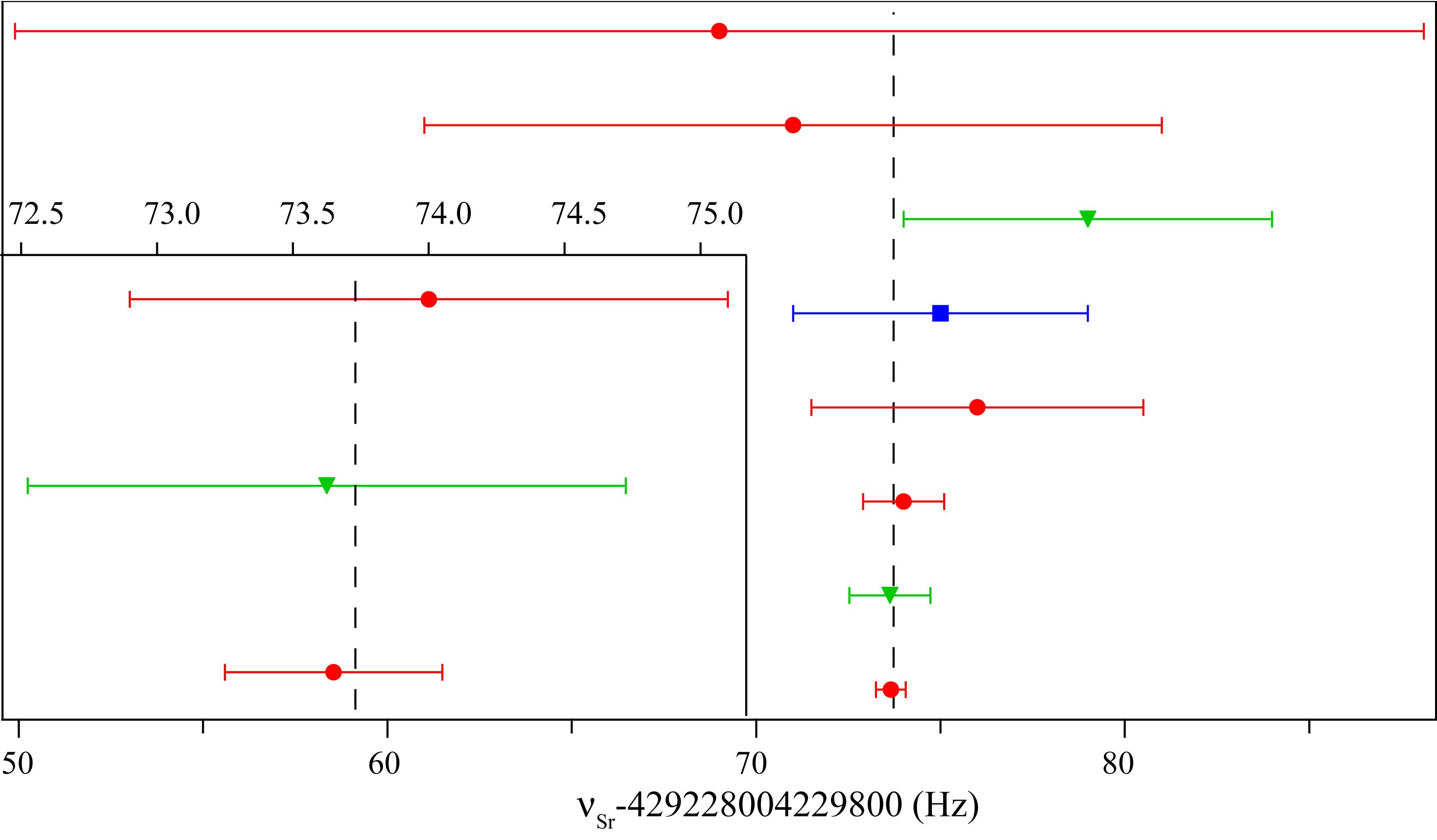}
\caption{\label{fig:figfive} Record of Sr absolute frequency
measurements. Previous measurements by this group (circles)
\cite{ludlow06,boyd06,ye06,boyd07}, as well as the Paris (triangle)
\cite{Targat06,Baillard07} and Tokyo (square) \cite{takamoto06}
groups are shown. The inset shows the high agreement of the most
recent measurements which agree below the $10^{-15}$ level. The
dashed line shows the weighted mean, $\bar{\nu}$ = 429 228 004 229
873.73 Hz of the combined data set. }}
\end{figure}

Including the uncertainty of the H-maser as well as the strontium
systematics described in Section III gives a final frequency of 429
228 004 229 873.65 (37) Hz, where the dominant uncertainty is due to
the H-maser calibration. In Fig.~\ref{fig:figfive}, this measurement
is compared with previous Sr frequency measurements by this group
\cite{ludlow06,boyd06,ye06,boyd07}, as well as by the Paris
\cite{Targat06,Baillard07} and Tokyo \cite{takamoto06} groups. As
shown in the figure, the agreement between international
measurements of the Sr frequency is excellent, with the most recent
measurements in agreement below the 10$^{-15}$ level, making the Sr
clock transition the best agreed upon optical frequency standard to
date. The high level of agreement enabled a recent analysis of this
combined data set that constrains the coupling of fundamental
constants to the gravitational potential as well as their drifts
\cite{Blatt08}.

\section{Conclusion}
In conclusion, we have made an accurate measurement of the
$^1S_0-$$^3P_0$ clock transition in fermionic strontium, where the
final fractional uncertainty of 8.6$\times$10$^{-16}$ is limited
primarily by the performance of the intermediate hydrogen maser used
to compare the Sr standard to the NIST-F1 Cs fountain clock. This
experiment represents one of the most accurate measurements of an
optical frequency to date, and the excellent agreement with previous
measurements makes strontium an excellent candidate for a possible
redefinition of the SI second in the future. In addition, the
combined frequency measurements of $^{87}$Sr performed worldwide, as
well as future measurements of frequency ratios with other optical
standards can be used to search for time-dependent frequency changes
which constrain variations of fundamental constants \cite{Blatt08}.

\section{Acknowledgement}
We gratefully thank S. Foreman and D. Hudson for their contribution
to the noise-cancelled fiber network, T. Fortier, J. Stalnaker, Z.
W. Barber, and C. W. Oates for the Sr - Ca optical comparison, and
J. Levine for help with the Cs-Sr elevation difference. We
acknowledge funding support from NIST, NSF, ONR, and DARPA. G.
Campbell is supported by a National Research Council postdoctoral
fellowship, M. Miranda is supported by a CAPES/Fullbright
scholarship, and J. W. Thomsen is a JILA visiting fellow, his
permanent address is The Niels Bohr Institute, Universitetsparken 5,
2100 Copenhagen, Denmark.

\bibliography{absolutefreq}

\begin{thebibliography}{56}
\bibitem[{geo()}]{geoid}
\bibinfo{note}{The geoid correction was determined using National Geodetic
  Survey benchmarks located on both the University of Colorado campus and the
  NIST campus, which have all been corrected using the GEOID03 model. Using
  these benchmarks, the geoid height between NIST and JILA varies by less than
  0.10 m.}
\end{thebibliography}

\end{document}